# 加工路径规划

邹强

浙江大学CAD&CG国家重点实验室

加工路径指工件与机床刀具的相对运动轨迹，其决定着刀具如何切削毛坯，以形成设计几何所规定的曲面形状。加工路径在很大程度上决定着进给速度、切屑面积、切削力甚至是颤振（Chatter）的发生与否,因此，路径质量直接关系着实际加工的精度和效率[1]。从CAD/CAM一体化角度来看，加工路径是CAD模型（即设计几何）和CNC加工代码的中间状态（即制造几何），是数字空间和物理空间之间的桥梁（见图一）。故而能否自动规划加工路径直接关系着产品生命周期的整体自动化程度和效率。

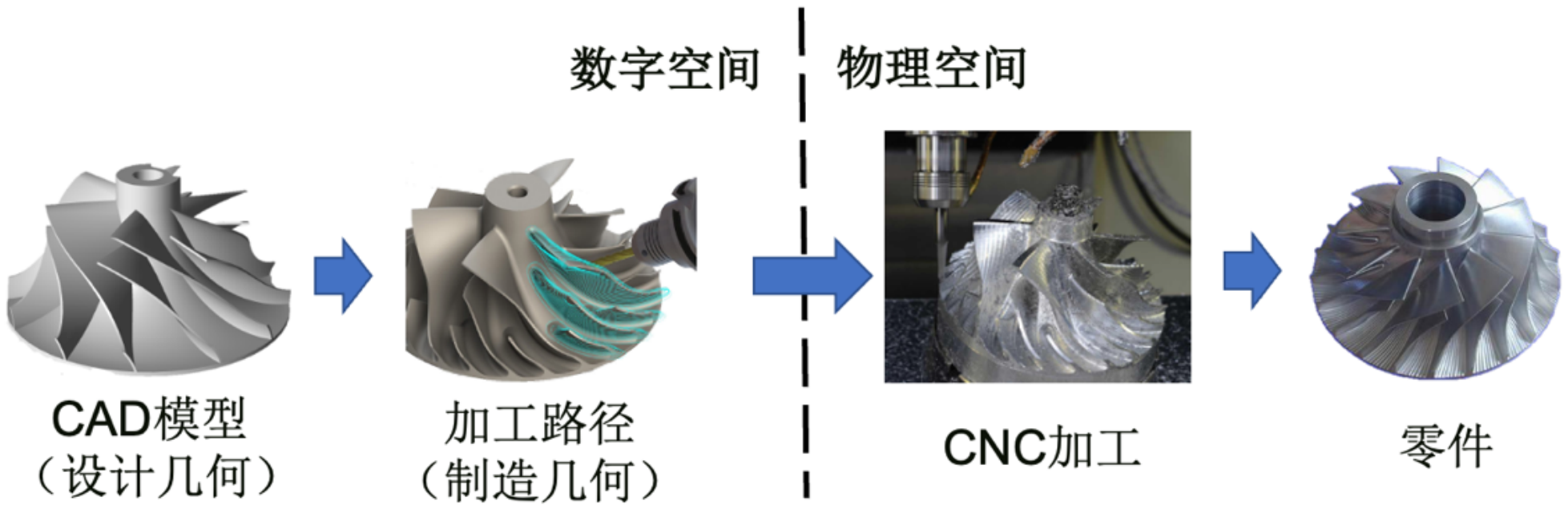

图一 加工路径是数字-物理二元空间的桥梁.

广义上，加工路径包括刀位点位置与刀轴方向两个方面；狭义上，加工路径仅指刀位点位置（一方面由于刀轴方向规划一般是在刀位点规划的基础之上进行，另一方面是因为3轴数控机床的广泛应用，这类机床的刀轴方向是固定不变的，不需要额外的刀轴方向规划）。此处，我们采用后者定义。加工路径的质量有三个主要评价标准[2－6]：

（1） 相邻路径间残留材料的高度（一般称为残高）不超过给定容差，以保证加工精度；

（2） 加工路径总长度尽量小，以减少加工时间；

（3） 加工路径的光滑程度尽量高，以防止频繁加减速，从而提高平均进给速度，降低加工时间。

其中，（1）是加工精度方面的要求，（2）和（3）是加工效率方面的要求。对于简单几何，如平面，这三个要求可以同时满足；然而，对于复杂曲面，他们之间往往是不相容的，不能同时满足。加工路径规划的难点就在于如何在这三者之间取得平衡，尤其是全局最优的平衡。

针对上述三个要求，人们在过去的30余年里提出了一系列路径规划方法，其中具有基础意义的加工路径生成方法主要有等参数法（iso-parametric）、等平面法（iso-plannar）、等残高法（iso-scallop）等[7]，新近发展的加工路径优化方法有场方法、等水平法等，分述如下。

## 一. 基础加工路径规划方法

在三大传统方法中，等参数法最早被学术界提出。如图二所示，对于三维空间中的自由曲面$S(u,v)$,该方法通过固定一个参数（比如参数$v$），变动另一参数（即参数$u$），在参数域生成一条直线段，并选取其所对应的三维空间曲线为加工路径[8]。相邻加工路径对应的参数增量$\Delta u$由工程师指定的容差确定[9]，即残高不超过给定容差值。

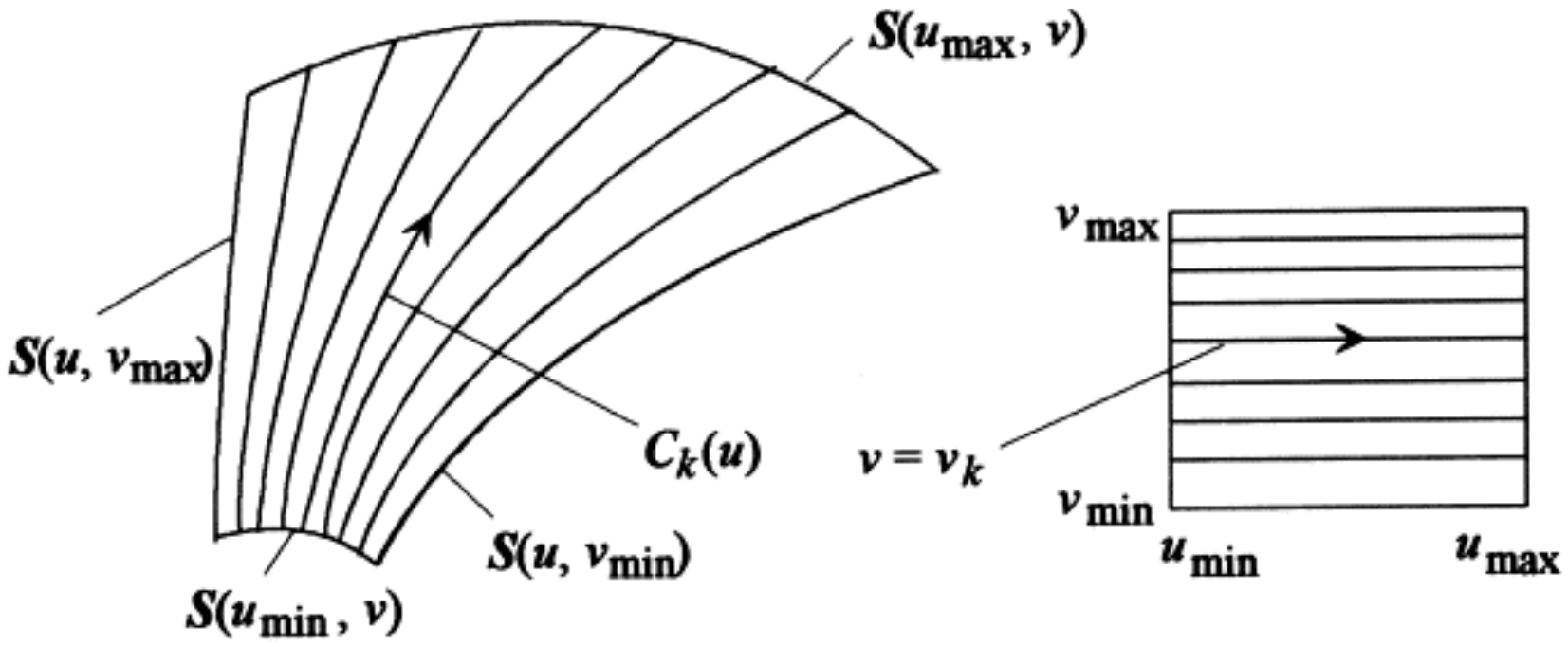

图二 等参数路径规划方法示意图[8].

等参数法具有路径光滑、计算简单、与NURBS曲面表示方法契合度高等优点，但也存在路径质量不高的缺点，主要表现为路径总长度过大（其原因在于路径疏密不一致、冗余加工过多），并且该方法不适用于离散的网格和点云曲面。针对前一缺点，常用的改进方法有：（1）对过密区域进行路径裁剪，例如[10]；（2）对过密区域进行路径间距微调，例如[11]。针对后一缺点，常用的解决方法是对网格模型和点云模型进行参数化，重建参数域[12]；这方面典型的工作有[13]和[14]，分别面向网格和点云。

通过上述改进方法，等参加工路径的总长度虽有所降低，但问题仍然突出。为进一步降低路径总长度，等平面法[1]被提出[15－18]。其基本思路是将待加工自由曲面与一组平行面相交，然后以相交线为加工路径，如图三所示。除使用一组平面外，人们还尝试过使用一组曲面与待加工自由曲面进行相交以生成加工路径，比如[19]（引导面方法中的投影操作在几何上等价于求交）。实际加工表明，等平面路径较等参路径具有更均匀的路径疏密度，更小的路径总长度[20]。与此同时，等平面法具有可直接应用于网格和点云模型的优势，比如[21]和[22]。然而，等平面路径仍未彻底解决路径疏密不一致、冗余加工过多的问题[23,24]。

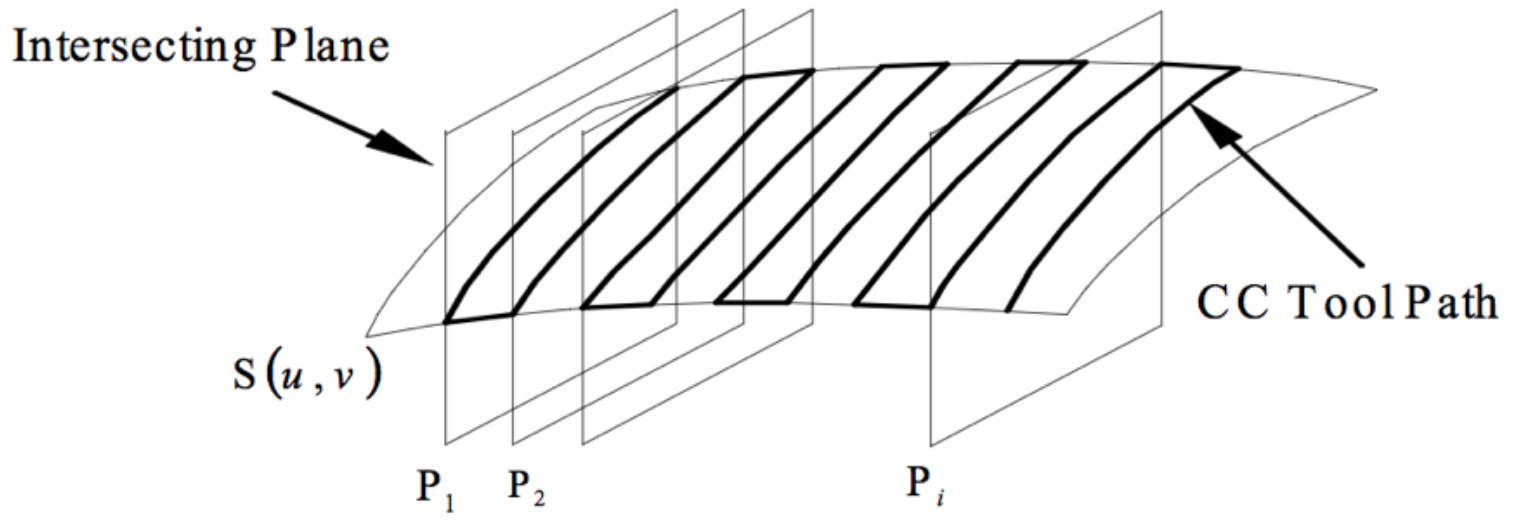

图三 等平面路径规划方法示意图[16].

对上述问题，常用的改进思路有两个：曲面分割法和等残高法。其中，曲面分割法将待加工曲面分为多个区域，然后在不同区域生成具有不同疏密度的等平面加工路径，以此来让各自区域内的路径疏密度尽量一致，这方面典型的工作有[16,25]。显然，这种方法只能在一定程度上减轻等平面法缺点带来的影响，不能完全解决问题，在各自区域内仍存在路径疏密不一致、冗余加工过多的问题。然而，曲面分割的思想在加工轨迹规划领域具有重要意义，最近这一思想又被重新挖掘出来，形成了近期基于方向场的加工轨迹规划方法[3,26]。

等残高法是一个能够完全解决路径疏密不一致和冗余加工问题的方法。其基本思路如图四所示，主要包括两部分[27]：（1）选定一条初始路径（如某段边界）并以此为基础在待加工曲面上生成一族偏移曲线；（2）在偏移时，相邻曲线间的残高保持恒定（即严格等于给定容差值，而不是像等参法或等平面法中的不超过给定容差值）。这一方法一经提出便广受关注，多个改进方案随后被提出，在提高计算精度方面典型的工作有[28－30]，在提高计算效率方面有[31－33]，

[1] 其实在等平面法被学术界正式提出之前，工业界已独立应用该方法[15,16]，但具体的算法步骤要么未披露要么不够系统。

在扩大应用范围方面（如平底刀具，网格或点云曲面）有[34 - 38]。与此同时，针对不同的加工需求，人们还将之拓展到如图五所示的多种拓扑形式[39,40]，例如螺旋形加工轨迹适用于高速加工。

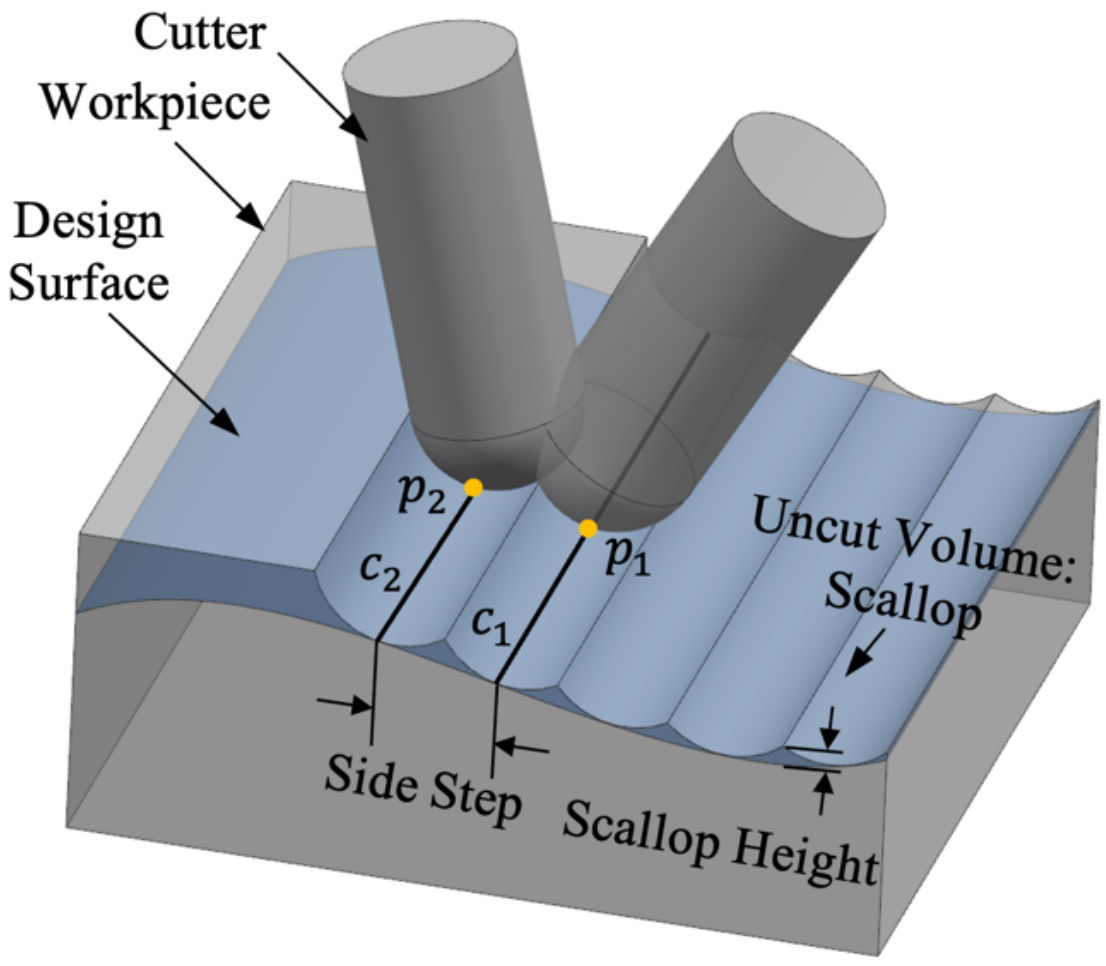

图四 等残高路径规划方法示意图[41].

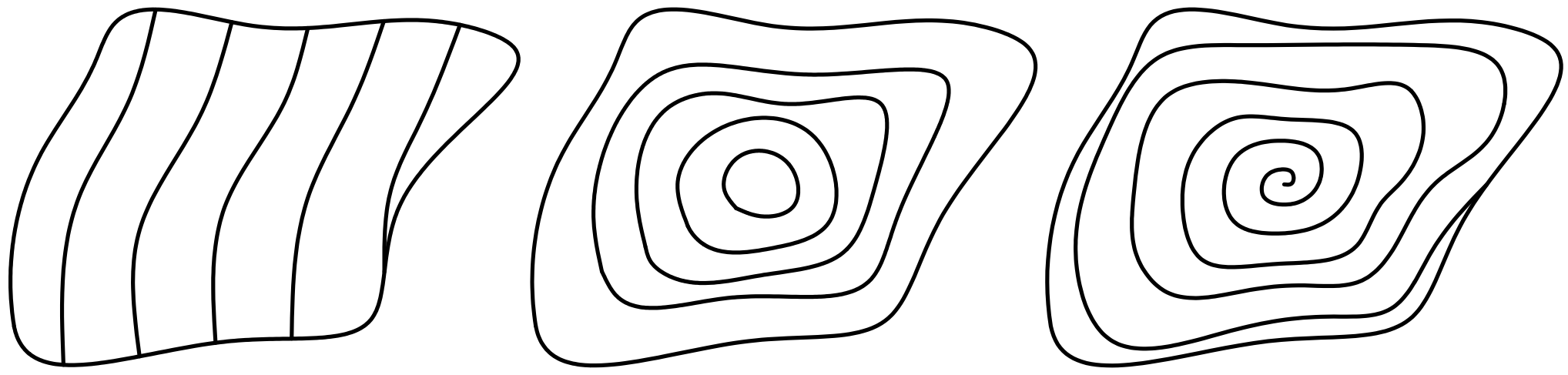

图五 加工路径拓扑：行型（左），回型（中）以及螺旋型（右）[14].

等残高法因彻底解决了冗余加工问题，而具有很小的加工路径总长度。然而，其也存在一定的问题，主要有以下两方面：

(1) 加工路径的总体形状依赖于初始路径的选取，其总长度也会有所不同，因此，等残高路径在总长度方面还有改进空间；

(2) 等残高加工路径的生成依赖于偏移操作，而曲线偏移往往会在路径上产生尖角（见图六），因此等残高路径在光滑度方面还有优化空间。

针对这两个问题，人们发展了一系列优化方法，例如最近的基于方向场的加工路径生成方法[3]便是一种通过方向场的流线来对路径进行优化选择的方法。

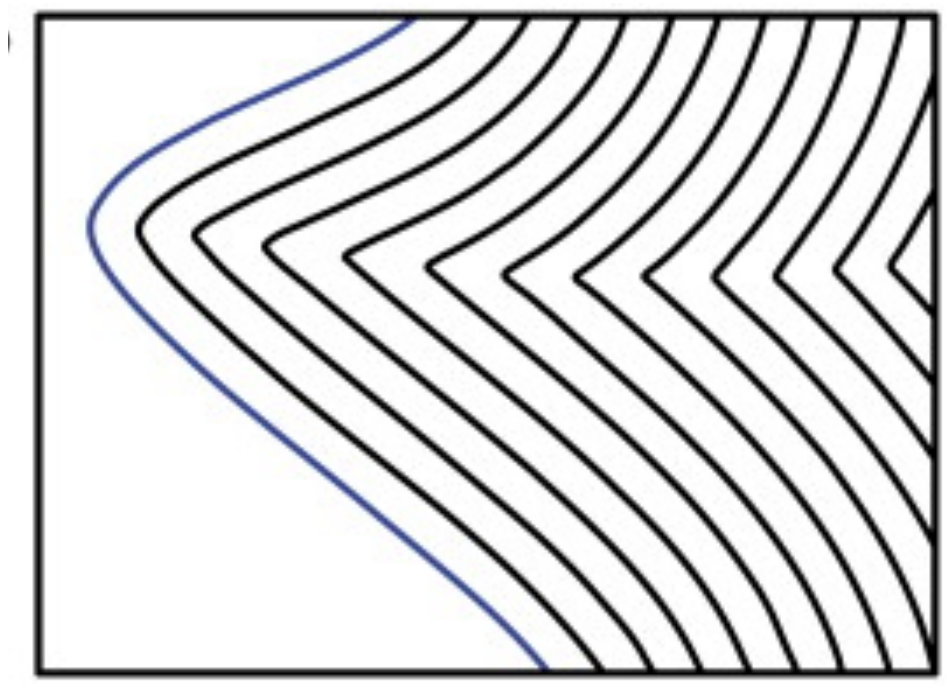

图六 曲线偏移导致尖角问题（蓝色为初始路径，黑色为偏移曲线）.

## 二．加工路径优化方法

针对初始路径选择的问题，初期的方法使用一种类似贪心算法的策略——所选取的初始路径具有最优的加工条件，其它路径仍然使用传统偏移的方法得到。最优初始路径的选取标准也多种多样，经典的有最大加工带宽（见图七）[6]、最大/最小曲面曲率[42]、最陡切向[43]等。这类贪心策略有一个明显的缺点，远离初始路径的加工路径的最优性难以保证。

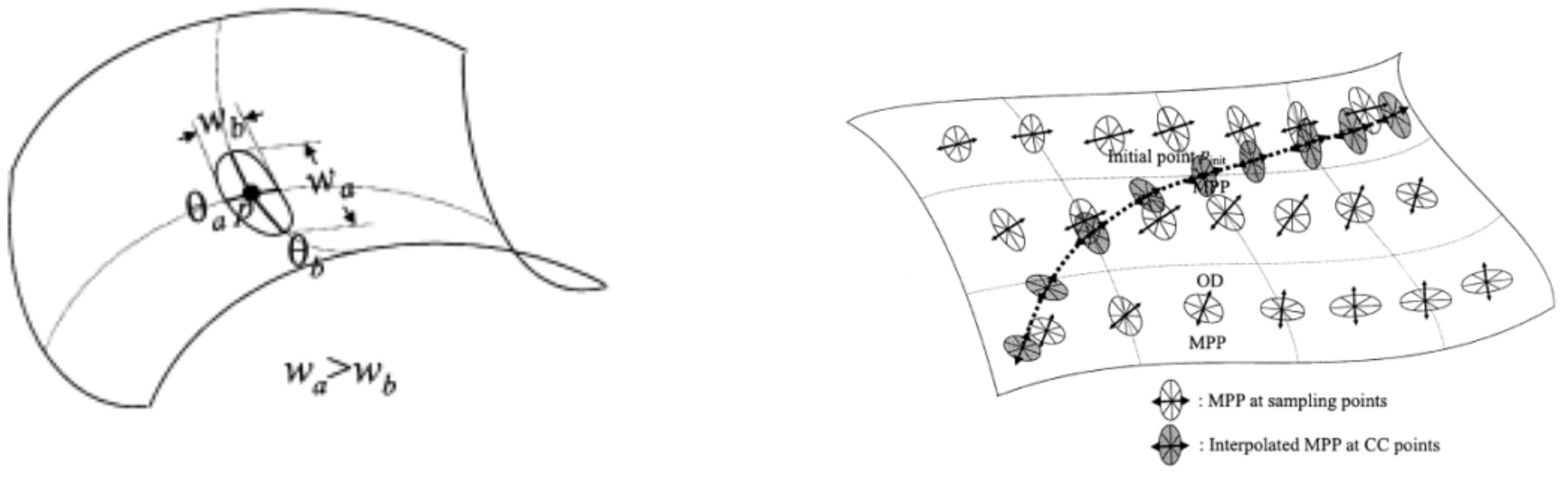

(a) (b)

图七 最大加工带宽示意图：(a)最大带宽方向示意图；(b)根据向量场生成最优初始路径.

上述问题的一大改进策略是对曲面进行分割，然后在各个被分割区域分别选取最优初始路径。由于被分割区域一般不大，初始路径与区域内的其它路径相距不远，他们的最优性在一定程度上能够被保证。传统上曲面分割方法是直接对曲率、加工带宽、运动学性能等加工性能指标进行自底向上的聚类或自顶向下的分类[16,42,44－46]，最近的曲面分割的方法是通过将加工性能指标转化成方向场

[2]，并利用场中的奇异点来对曲面进行分割，如图八和图九所示。场方法的依据在于奇异点分割出来的区域内部的流线一般具有很好的相似性[47]，这可以保持偏移路径与初始路径的一致性。

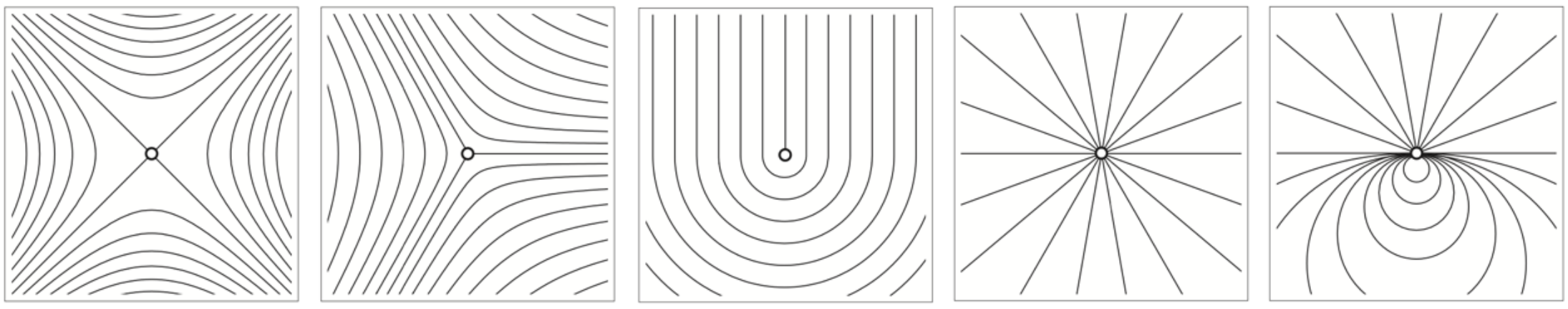

图八 方向场奇异点示意图.

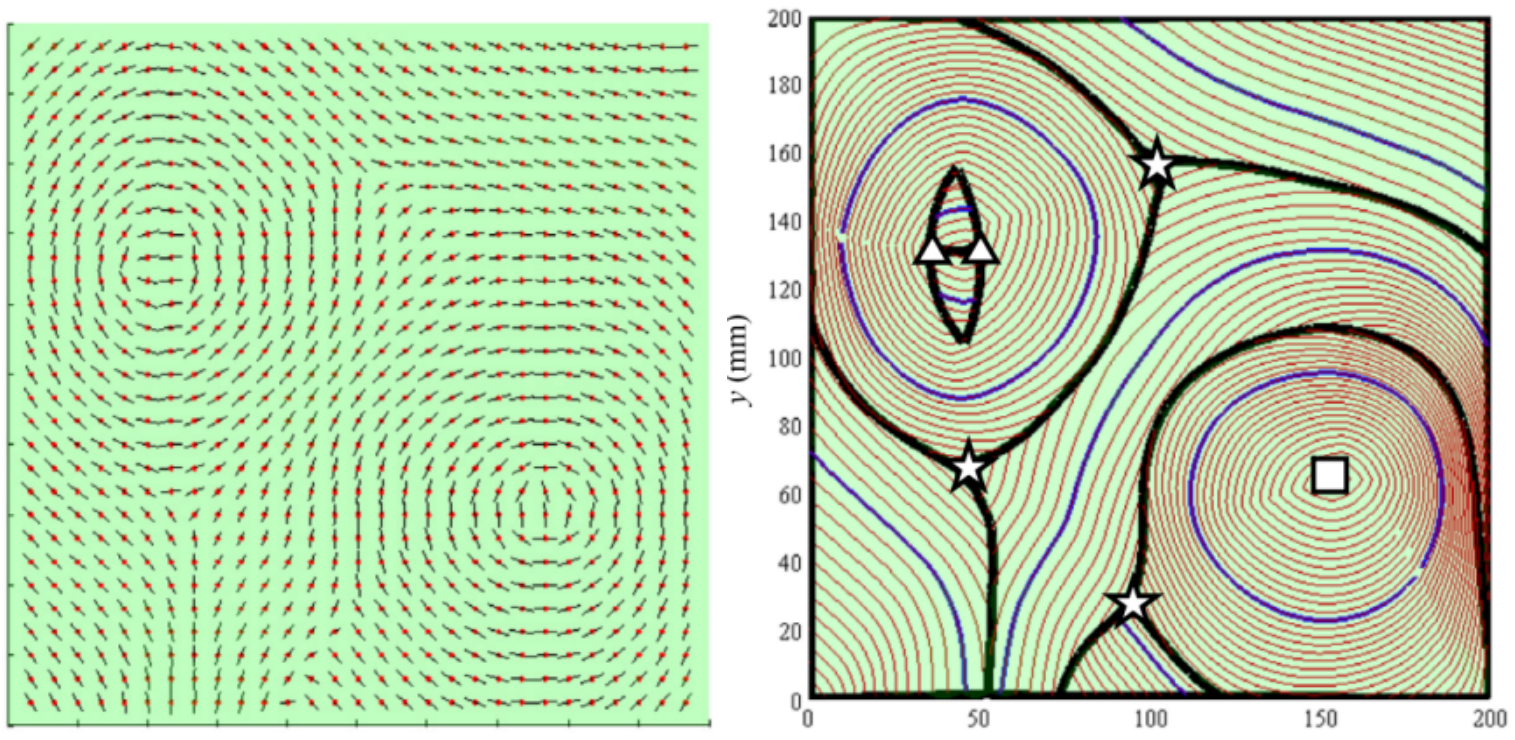

(a) (b)

图九 基于方向场的曲面分割方法：(a)代表最优加工方向的方向场；(b)根据奇异分界对曲面进行分割.

利用方向场（一说张量场，本质一样）的奇异点来进行曲面分割这一思想首先由 Kumazawa 和 Feng 提出[3,26]。随后，一系列改进方法被提出[48-60]，主要沿着以下三个方面展开研究：

(1) 向量场生成方法：将生成方向场的加工性能指标从最大加工带宽推广至最大材料去除率[55]、最大进给速度[59]、能量消耗最小[60]或组合指标[52,58]；

(2) 分割区域内路径生成方法：将[3,26]中的初始路径人工选择方法扩展至自动选择方法[57]；将[3,26]中的等残高路径改为纯流线的路径[51-53,56]（这一改变实质上反而增加路径长度）；

(3) 分割区域内路径拓扑确定方法：将[3,26]中的人工指定路径拓扑方法扩

---

[2] 很多文献中使用矢量场来描述这类方法，这是不正确的，此类方法只利用了方向信息，不利用模长信息。

展至根据流线形状自动确定路径拓扑的方法（见图十）[48,49]。

虽然基于方向场的加工路径规划方法取得了很好的效果，方向场也具有表示多种加工指标的优点，但仍有一个重要问题有待回答：按奇异点来分割曲面是否是全局最优分割，分割区域内路径生成方法是否为全局最优？目前的方法在这两方面都是采用启发式算法，其与全局最优路径的关系尚未知。最近的[50]提出了一种将方向场推广至矢量场的方法，首次实现了路径总长度的全局优化，但尚未包含其它加工指标的全局优化。总而言之，场方法未来仍有很大的改进空间。

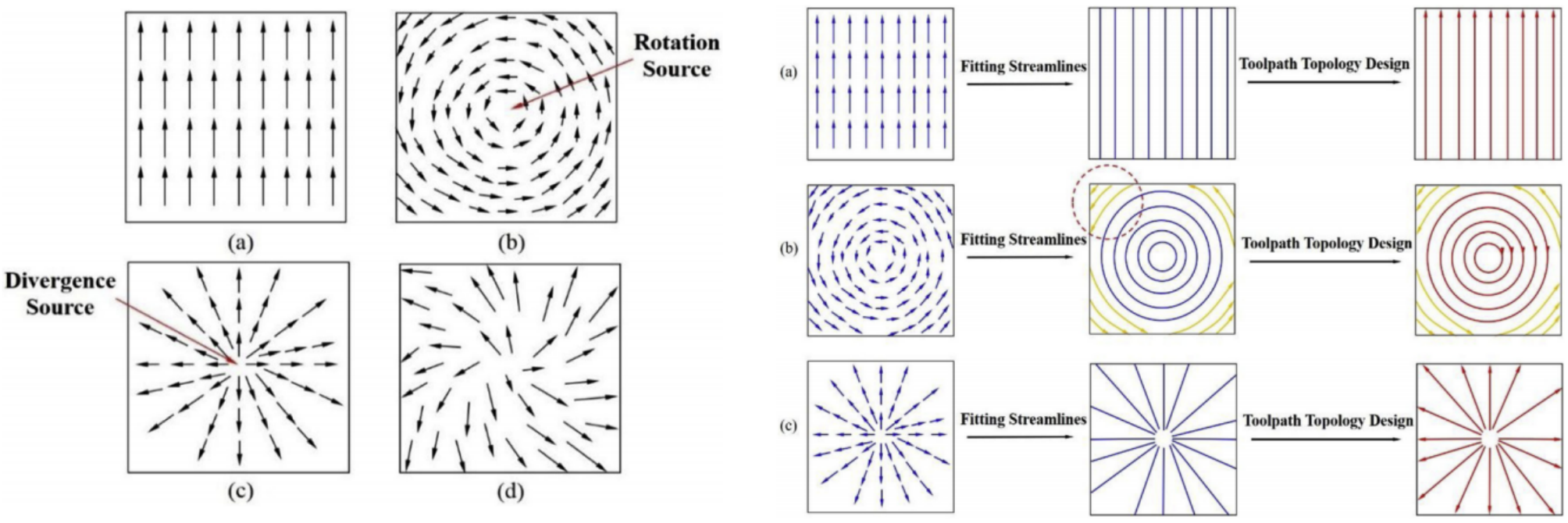

(a) (b)

图十 方法[49]示意图：(a)根据向量场对待加工区域分类；(b)不同类别设计不同拓扑的路径.

与路径长度优化相比，有关路径光滑度的工作较少。初期的方法是在已有路径基础上对尖角部分进行局部修正[61－63]。这类方法具有实现简单的优点，但也存在两个问题：其一，在修正处破坏了等残高性质；其二，局部修正只能给出次优路径，不是全局最优路径。

近期提出的等水平方法是一种加工路径全局优化方法。与传统方法不同，其采用隐式加工路径表示方法（见图十一），并将各加工指标（如等残高，最大带宽）表达为能量函数，然后通过最小化这些能量函数获得全局最优加工路径[5]。这一隐式优化框架的后续改进工作已使其能够全局优化残高分布[41]、光滑度[41,51]、长度[50]、进给速度[64]等。这一方法与场方法的结合具有很大的发展潜力。

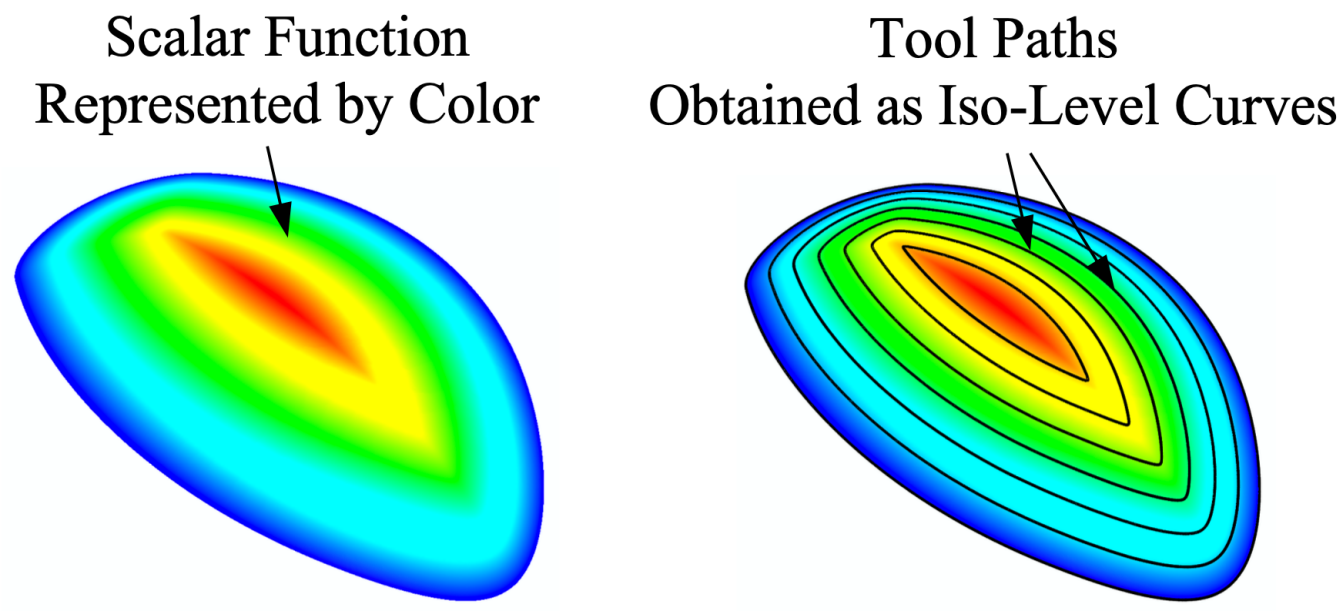

图十一 加工路径隐式表示示意图[41].

除上述方法外，人们还将启发式（MetaHeuristic）算法，如基因算法、粒子群算法，引入到加工路径全局优化中，代表性工作有[65 - 67]。这部分工作虽然仍处于初期，但是具有很大的发展潜力。

**参考文献：**